\documentclass[reprint, twocolumn,
 amsmath,amssymb,
 aps, floatfix,
]{revtex4-2}

\usepackage{graphicx}
\usepackage{dcolumn}
\usepackage{bm}
\usepackage[utf8]{inputenc}
\usepackage[T1]{fontenc}

\begin{document}

\preprint{APS/123-QED}

\title{\textbf{Differential Geometry of Deterministic Stokesian Dynamics}}

\author{Sumedh R. Risbud}
\affiliation{Principal Contributor, Mathematical Modeling Initiative \\
 Pune, India.}\altaffiliation[{\normalsize Also at }]{{\normalsize Neuromorphic Computing Group, Intel}}\altaffiliation[{\normalsize and }]{{\normalsize Department of Electrical Engineering, IIT Bombay}}
\email{\\{\tt srrisbud@proton.me}}
\homepage{\\{\tt http://www.linkedin.com/in/sumedhratnakarrisbud}}

\date{\today}

\begin{abstract}
We present a differential geometric formulation of deterministic, low-Reynolds-number particle dynamics in quiescent fluids with fixed obstacles. Within this framework, we discover and prove a new trajectory-level extremum principle for Stokesian suspensions: the physical trajectory of a force-driven particle extremizes the cumulative energy dissipated into the fluid, corresponding to an exact geodesic of a conformally scaled Riemannian manifold, $\tilde{g}_{ij} = \mathcal{D}(\mathbf{x})\tilde{R}_{ij}$, where $\tilde{\mathbf{R}}$ is the effective resistance tensor and $\mathcal{D}$ is the instantaneous power dissipation. While the Helmholtz minimum dissipation theorem governs the Eulerian fluid velocity field at each instantaneous snapshot, it cannot directly predict how an externally forced particle selects its path over time. One might expect the hydrodynamic resistance tensor to serve directly as a metric, but we show that spatial resistance gradients cause particle trajectories to drift away from bare resistance geodesics. This drift is resolved by conformally scaling the effective resistance tensor $\tilde{\mathbf{R}}$ by the instantaneous power dissipation $\mathcal{D}(\mathbf{x})$, establishing a geometric principle valid for arbitrary obstacle arrangements. In this geometry, physical laboratory time is non-affine, and the affine parameter measuring trajectory length is the cumulative energy dissipated into the fluid. We validate this principle against exact analytical solutions for a sphere traversing past a fixed spherical obstacle and outline its application to designing microfluidic particle sorting geometries.
\end{abstract}

\maketitle

\section{Introduction}
The use of geometry to describe physical phenomena is a cornerstone of modern theoretical physics, most notably in Einstein's description of gravitation \cite{Einstein1915} and the broader field of geometric mechanics \cite{Arnold1966, HolmSchmahStoica2009}. In these classical frameworks, the physical system is typically conservative, and the logical progression is straightforward: a Lagrangian is defined, the Principle of Least Action yields the equations of motion, and these are interpreted as geodesic equations on a Riemannian manifold.

Here, we extend this geometric treatment to the purely dissipative, Aristotelian domain of low-Reynolds-number hydrodynamics. The Stokesian regime, defined by negligible fluid and particle inertia, is governed by the linear Stokes equations:
\begin{align}\label{eqn:stokeseq}
    \mu \nabla^2 \mathbf{u} &= \nabla p, \quad \nabla\cdot\mathbf{u} = 0.
\end{align}
Because the equations are linear and time-reversible, the motion of a suspended rigid particle is governed by the $6 \times 6$ Grand Resistance Tensor $\bm{\mathcal{R}}$ relating the hydrodynamic force $\mathbf{F}$ and torque $\mathbf{T}$ to the translational velocity $\mathbf{U}$ and angular velocity $\boldsymbol{\omega}$ \cite{HappelBrenner, KimKarrila}:
\begin{equation}
    \begin{pmatrix} \mathbf{F} \\ \mathbf{T} \end{pmatrix} = 
    \begin{pmatrix} \mathbf{R}_{\mathbf{FU}}(\mathbf{x}) & \mathbf{R}_{\mathbf{F\omega}}(\mathbf{x}) \\ \mathbf{R}_{\mathbf{TU}}(\mathbf{x}) & \mathbf{R}_{\mathbf{T\omega}}(\mathbf{x}) \end{pmatrix} 
    \begin{pmatrix} \mathbf{U} \\ \boldsymbol{\omega} \end{pmatrix}.
    \label{eqn:grand_resistance}
\end{equation}
Here, the coupling resistance tensor $\mathbf{R}_{\mathbf{F\omega}}$ is the transpose of $\mathbf{R}_{\mathbf{TU}}$ (i.e., $\mathbf{R}_{\mathbf{F\omega}} = \mathbf{R}_{\mathbf{TU}}^T$) by virtue of the Lorentz reciprocal theorem. The $3 \times 3$ sub-matrices $\mathbf{R}_{\mathbf{FU}}(\mathbf{x})$, $\mathbf{R}_{\mathbf{F\omega}}(\mathbf{x})$, and $\mathbf{R}_{\mathbf{T\omega}}(\mathbf{x})$ represent translational resistance, translation-rotation coupling, and rotational resistance, respectively. In the presence of obstacles or boundaries (as well as for non-orthotropic particles), the coupling tensor $\mathbf{R}_{\mathbf{F\omega}}$ generally does not vanish, causing translating particles to rotate.

\subsection{Effective Resistance for Torque-Free Particles}
For a freely suspended particle driven by a constant net external force $\mathbf{F}$ (such as gravity minus buoyancy) applied through its center of force, no external torque acts on the body ($\mathbf{T} = \mathbf{0}$). Setting $\mathbf{T}=\mathbf{0}$ in Eq.~\eqref{eqn:grand_resistance} determines the slaved rotational velocity \cite{HappelBrenner, KimKarrila}:
\begin{equation}
    \boldsymbol{\omega} = -\mathbf{R}_{\mathbf{T\omega}}^{-1}\mathbf{R}_{\mathbf{F\omega}}^T\mathbf{U}.
    \label{eqn:forced_rotation}
\end{equation}
Substituting Eq.~\eqref{eqn:forced_rotation} into the translational force balance yields the effective resistance relation:
\begin{equation}
    \mathbf{F} = \tilde{\mathbf{R}}(\mathbf{x}) \cdot \mathbf{U},
    \label{eqn:linear_eff_res}
\end{equation}
where $\tilde{\mathbf{R}}(\mathbf{x})$ is the Schur complement of the rotational block:
\begin{equation}
    \tilde{\mathbf{R}}(\mathbf{x}) \equiv \mathbf{R}_{\mathbf{FU}}(\mathbf{x}) - \mathbf{R}_{\mathbf{F\omega}}(\mathbf{x})\mathbf{R}_{\mathbf{T\omega}}^{-1}(\mathbf{x})\mathbf{R}_{\mathbf{F\omega}}^T(\mathbf{x}).
    \label{eqn:Rtilde_def}
\end{equation}
The instantaneous rate of power dissipation into the fluid is then:
\begin{equation}
    \mathcal{D} = \mathbf{F}\cdot\mathbf{U} = \mathbf{U}\cdot\tilde{\mathbf{R}}\cdot\mathbf{U}.
    \label{eqn:dissip_coupling}
\end{equation}
The tensor $\tilde{\mathbf{R}}(\mathbf{x})$ defines the natural base metric on the translational configuration space of particle positions.

\subsection{Helmholtz Dissipation and Particle Trajectories}
Variational descriptions of Stokes flow originate with the Helmholtz minimum dissipation theorem \cite{Helmholtz1868}, later extended by Hill and Power \cite{HillPower1956} and Keller et al. \cite{KellerRubenfeldMolyneux1967} to include suspended particles. The theorem states that among all solenoidal velocity fields satisfying prescribed boundary velocities, the Stokes flow field minimizes the instantaneous volume dissipation rate.

However, the Helmholtz theorem is an instantaneous Eulerian principle. For a fixed particle position $\mathbf{x}$, the fluid flow minimizes dissipation for given boundary velocities, but the theorem does not determine how a particle driven by an external force selects its trajectory $\mathbf{x}(t)$. Under a constant driving force, the velocity is not prescribed; rather, it adjusts locally via $\mathbf{U}(\mathbf{x}) = \tilde{\mathbf{R}}^{-1}(\mathbf{x})\cdot\mathbf{F}$.

This raises the question of whether a trajectory-level variational principle governs the motion of Stokesian particles through obstacle fields. One might examine whether particle paths follow geodesics of the effective resistance metric $\tilde{\mathbf{R}}$, which would correspond to paths minimizing cumulative resistance.

Below, we show that particle trajectories driven by constant forces deviate from the geodesics of $\tilde{\mathbf{R}}$ due to a geometric drift arising from spatial gradients in the resistance tensor. Instead, physical trajectories are exact geodesics of a conformally scaled metric, $\tilde{\mathbf{g}} = \mathcal{D}\tilde{\mathbf{R}}$, where $\mathcal{D}$ is the instantaneous power dissipation. In this metric space, the affine parameter is the cumulative energy dissipated into the fluid.

\section{Particle trajectories and Geodesics}\label{sec:synthesis}
We consider the energy dissipated along a particle trajectory:
\begin{equation}
    E = \int_0^{t} \tilde{\mathbf{R}}\cdot\mathbf{U}\cdot\mathbf{U}~d\tau, \label{eqn:energyintegralquadform}
\end{equation}
where the integral runs over the trajectory from $\tau=0$ to $\tau=t$.

Because the instantaneous velocity $\mathbf{U}$ is the tangent vector to the path, Eq.~\eqref{eqn:energyintegralquadform} can be written in index notation as:
\begin{equation}
E = \int_0^{t} \tilde{R}_{ij}\frac{dx^i}{d\tau}\frac{dx^j}{d\tau}~d\tau, \label{eqn:energyintegralindex}
\end{equation}
using the summation convention on a manifold endowed with the metric $g_{ij} = \tilde{R}_{ij}$. Lower indices denote covariant components, upper indices denote contravariant components, and $g^{ij}$ denotes the matrix inverse of $g_{ij}$.

Applying the Euler-Lagrange equations to Eq.~\eqref{eqn:energyintegralindex} yields the geodesic equations for the metric $g_{ij} = \tilde{R}_{ij}$. We now test whether physical trajectories driven by a constant external force $F_k$ satisfy these equations. The equation of motion is:
\[U^i = (g_{ij})^{-1}F_j = g^{ij} F_j = a^{ij} F_j,\] 
where $g^{ij} = (\tilde{R}_{ij})^{-1} = a^{ij}$ is the contravariant mobility tensor.
Differentiating with respect to time yields the kinematic acceleration:
\begin{equation}
    \frac{dU^i}{dt} = \frac{\partial g^{ij}}{\partial x^k} U^k F_j.
\end{equation}
Using the identity $\partial_k g^{ij} = -g^{ia} g^{jb} (\partial_k g_{ab})$ and substituting $F_j = g_{jl} U^l = \tilde{R}_{jl} U^l$, we obtain:
\begin{align}
    \frac{dU^i}{dt}
    &= -g^{ia}g^{jb}(\partial_k g_{ab})U^kF_j \nonumber\\
    &= -g^{ia}g^{jb}(\partial_k g_{ab})U^k g_{jl}U^l \nonumber\\
    &= -g^{ia}(\partial_k g_{al})U^kU^l.
\end{align}
The covariant derivative along the curve is $\frac{DU^i}{dt} = \frac{dU^i}{dt} + \Gamma^i_{jk} U^j U^k$, where the Christoffel symbols for the metric $g_{ij} \equiv \tilde{R}_{ij}$ are:
\begin{equation}
    \Gamma^i_{jk} = \frac{1}{2}g^{im}(\partial_j g_{mk} + \partial_k g_{mj} - \partial_m g_{jk}).
\end{equation}
Substituting the kinematic acceleration yields:
\begin{align}
    \frac{DU^i}{dt}
    &= -g^{im}(\partial_k g_{mj})U^kU^j \nonumber \\ 
    &~~~ + \frac{1}{2}g^{im}(\partial_j g_{mk}+\partial_k g_{mj}-\partial_m g_{jk})U^jU^k \nonumber\\
    &= -\frac{1}{2}g^{im}(\partial_m g_{jk})U^jU^k,
\end{align}
where dummy indices have been relabeled using the symmetry $U^jU^k = U^kU^j$.

\begin{figure}[t]
    \centering
    \includegraphics[width=\linewidth]{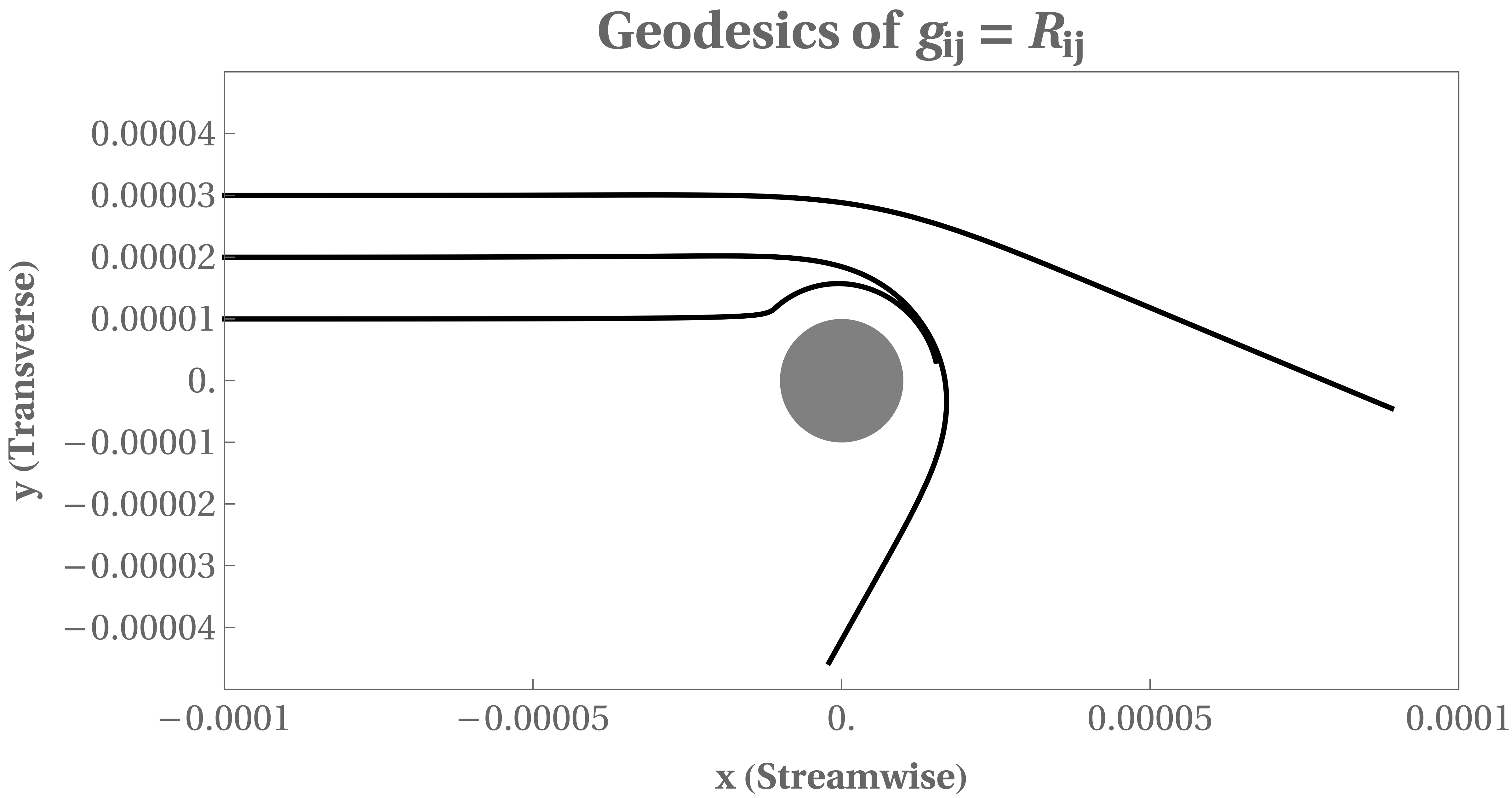}
    \caption{Geodesics of the bare effective resistance metric $g_{ij} = \tilde{R}_{ij}$. The fixed spherical obstacle is shown in gray. The particle and obstacle have radii $a = 10~\mu$m in water ($\mu=10^{-3}$ Pa$\cdot$s, $\rho_f=1000$ kg/m$^3$, particle density $2000$ kg/m$^3$). Impact parameters are $\{a, 2a, 3a\}$ from bottom to top.}
    \label{fig:gIsRTraj}
\end{figure}

Using the instantaneous power dissipation $\mathcal{D} = g_{jk} U^j U^k = \tilde{R}_{jk} U^j U^k$, we have:
\begin{align}
    \partial_m\mathcal{D} &= F_j\partial_m U^j \nonumber \\
    &= F_j\partial_m(g^{jk}F_k) \nonumber \\
    &= F_j(\partial_m g^{jk})F_k \nonumber \\
    &= -g_{ja}U^a g^{jb} (\partial_m g_{bc}) g^{kc} g_{kl} U^l \nonumber \\
    &= -\partial_m g_{jk} U^j U^k.
\end{align}
The covariant acceleration in the effective resistance metric is therefore:
\begin{equation} \label{eqn:geodesic_deviation}
    \frac{DU^i}{dt} = \frac{1}{2}g^{im}\partial_m\mathcal{D} \neq 0.
\end{equation}

\subsection{Geodesic Deviation and Fore-Aft Symmetry}
Equation~\eqref{eqn:geodesic_deviation} shows that the particle trajectory is not a geodesic of the bare resistance metric alone. The non-zero covariant acceleration $\frac{DU^i}{dt} = \frac{1}{2}g^{im}\partial_m\mathcal{D}$ indicates a geometric drift resulting from spatial gradients in the dissipation rate.

For a two-sphere scattering problem (detailed in \S\ref{sec:two-sphere-system}), Figure~\ref{fig:gIsRTraj} shows the numerical geodesics obtained by integrating $\frac{DU^i}{dt} = 0$ for $g_{ij} = \tilde{R}_{ij}$. These paths deflect outward upon approaching the obstacle and continue to diverge downstream, failing to return to their undisturbed horizontal orientation.

This occurs because the metric $g_{ij} = \tilde{R}_{ij}$ measures only the local resistance of the medium. Near the obstacle, lubrication forces increase the resistance, so geodesics of $\tilde{\mathbf{R}}$ steer away from high-resistance regions. However, the linearity and time-reversibility of Stokes flow require the trajectory of a sphere moving past a fixed sphere under a constant external force to be symmetric about the obstacle midplane ($x = 0$). The bare resistance metric does not account for the work done by the driving force and cannot supply the restorative steering needed downstream. A particle driven by a constant force balances the resistance gradient against the local power dissipation.

\section{The Jacobi-Maupertuis Resolution}\label{sec:jacobi_maupertuis}
The trajectories in Figure~\ref{fig:gIsRTraj} deviate from physical paths because the metric $g_{ij} = \tilde{R}_{ij}$ accounts for fluid resistance but omits the energetics of the driving force. To connect the kinematics to a variational formulation, we introduce a dissipative analogue of the Jacobi-Maupertuis principle.

In classical mechanics, the Jacobi-Maupertuis metric $\tilde{g}_{ij} = 2(E-V)g_{ij}$ converts the motion of a conservative system at fixed energy $E$ into a geodesic problem on a conformally scaled manifold. In low-Reynolds-number flows, inertia is negligible, and mechanical work done by the external force is dissipated as viscous heat. The instantaneous rate of power dissipation along the trajectory is:
\begin{equation}
    \mathcal{D} = \mathbf{F}\cdot\mathbf{U} = \mathbf{U} \cdot \tilde{\mathbf{R}} \cdot \mathbf{U} = \tilde{R}_{ij} \dot{x}^i \dot{x}^j.
\end{equation}
Near an obstacle, hydrodynamic resistance increases and particle speed drops ($\mathbf{U} \propto \tilde{\mathbf{R}}^{-1}\mathbf{F}$), reducing the dissipation rate $\mathcal{D} = \mathbf{F}\cdot\mathbf{U}$.

We consider a conformally scaled metric:
\begin{equation} \label{eqn:dissipation_metric}
    \tilde{g}_{ij}(\mathbf{x}) = \mathcal{D}(\mathbf{x}) \tilde{R}_{ij}(\mathbf{x}).
\end{equation}
In this metric, the squared norm of the velocity vector is $\tilde{g}_{ij} \dot{x}^i \dot{x}^j = \mathcal{D} (\tilde{R}_{ij} \dot{x}^i \dot{x}^j) = \mathcal{D}^2$, so that $\mathcal{D} = \sqrt{\tilde{g}_{ij} \dot{x}^i \dot{x}^j}$. The action functional can thus be written as a Riemannian length functional:
\begin{equation}
    S = \int_0^T \mathcal{D}(\mathbf{x}, \dot{\mathbf{x}}) \, dt = \int_0^T \sqrt{\tilde{g}_{ij} \frac{dx^i}{dt} \frac{dx^j}{dt}} \, dt. \label{eqn:action_length}
\end{equation}
Curves that extremize this functional are geodesics of the metric $\tilde{\mathbf{g}}$.

\subsection{Geodesic Equations under Conformal Scaling}
To verify that physical particle trajectories are geodesics of this metric, consider a generic conformal scaling:
\begin{equation}
    \tilde{g}_{ij} = \Omega(\mathbf{x}) \tilde{R}_{ij},
\end{equation}
for an arbitrary spatial scaling function $\Omega(\mathbf{x})$. The Christoffel symbols transform as:
\begin{equation}
    \tilde{\Gamma}^i_{jk} = \Gamma^i_{jk} + \underbrace{\frac{1}{2\Omega} \left( \delta^i_k \partial_j \Omega + \delta^i_j \partial_k \Omega - g_{jk} g^{im} \partial_m \Omega \right)}_{C^i_{jk}}, 
\end{equation}
where $\Gamma^i_{jk}$ are the Christoffel symbols of the bare metric $g_{ij} \equiv \tilde{R}_{ij}$.
The covariant derivative of velocity in the scaled metric is:
\begin{equation}
    \frac{\tilde{D}U^i}{dt} = \frac{dU^i}{dt} + \tilde{\Gamma}^i_{jk} U^j U^k.
\end{equation}
Substituting $\tilde{\Gamma} = \Gamma + C$ gives:
\begin{align}
    \frac{\tilde{D}U^i}{dt}
    &= \underbrace{\left( \frac{dU^i}{dt} + \Gamma^i_{jk} U^j U^k \right)}_{DU^i/dt} \nonumber \\ 
    &~~+ \frac{1}{2\Omega} \left( 2 U^i (U^k \partial_k \Omega) - (g_{jk} U^j U^k) g^{im} \partial_m \Omega \right).
\end{align}
Using $U^k \partial_k \Omega = \dot{\Omega}$ and $g_{jk} U^j U^k = \mathcal{D}$, this simplifies to:
\begin{equation} \label{eqn:general_accel}
    \frac{\tilde{D}U^i}{dt} = \frac{DU^i}{dt} + \frac{\dot{\Omega}}{\Omega} U^i - \frac{\mathcal{D}}{2\Omega} g^{im} \partial_m \Omega.
\end{equation}
Substituting $\frac{DU^i}{dt} = \frac{1}{2} g^{im} \partial_m \mathcal{D}$ from Eq.~\eqref{eqn:geodesic_deviation} into Eq.~\eqref{eqn:general_accel} yields:
\begin{equation}
    \frac{\tilde{D}U^i}{dt} = \frac{1}{2} g^{im} \left( \partial_m \mathcal{D} - \frac{\mathcal{D}}{\Omega} \partial_m \Omega \right) + \frac{\dot{\Omega}}{\Omega} U^i.
\end{equation}
Choosing $\Omega(\mathbf{x}) = \mathcal{D}(\mathbf{x})$, the transverse acceleration term vanishes:
\begin{equation}
    \partial_m \mathcal{D} - \frac{\mathcal{D}}{\mathcal{D}} \partial_m \mathcal{D} = 0,
\end{equation}
leaving:
\begin{equation}
    \frac{\tilde{D}U^i}{dt} = \frac{\dot{\mathcal{D}}}{\mathcal{D}} U^i. \label{eqn:projective_geodesic}
\end{equation}
Equation~\eqref{eqn:projective_geodesic} is the projective geodesic equation: the covariant acceleration is parallel to the velocity vector, showing that the physical trajectory is an unparameterized geodesic of $(\mathcal{M}, \tilde{\mathbf{g}})$.

The conformal factor $\mathcal{D}(\mathbf{x})$ acts as a restorative term opposing the bare resistance gradient. Upstream of the obstacle ($x < 0$), the particle slows down as it approaches the obstacle ($\partial_x \mathcal{D} < 0$), counteracting the outward deflection. Downstream ($x > 0$), the particle accelerates toward its undisturbed speed, reversing the sign of $\nabla \mathcal{D}$. This gradient reversal steers the trajectory back toward the axis of symmetry, recovering the fore-aft symmetry of Stokes flow.

\subsection{Affine Parameter and Trajectory Extremum Principle}
Because $\frac{\tilde{D}U^i}{dt} = \frac{\dot{\mathcal{D}}}{\mathcal{D}} U^i$, laboratory time $t$ is non-affine. Introducing an affine parameter $s(t)$ with tangent vector $V^i = dx^i/ds$ such that $U^i = \dot{s} V^i$, the covariant acceleration expands as:
\begin{equation}
    \frac{\tilde{D}U^i}{dt} = \ddot{s} V^i + \dot{s}^2 \frac{\tilde{D}V^i}{ds}.
\end{equation}
Imposing the affine condition $\frac{\tilde{D}V^i}{ds} = 0$ yields:
\begin{align}
    \ddot{s} = \frac{\dot{\mathcal{D}}}{\mathcal{D}} \dot{s} &\implies \frac{d}{dt}(\ln \dot{s}) = \frac{d}{dt}(\ln \mathcal{D}) \nonumber \\ 
    & \implies \dot{s} = \mathcal{D} \nonumber \\ 
    & \implies ds = \mathcal{D} \, dt. \label{eqn:timedilationfinal}
\end{align}
Equivalently, defining $s$ via the metric arc length, $ds = \sqrt{\tilde{g}_{ij} dx^i dx^j} = \mathcal{D} dt$, the Euler-Lagrange equations yield the standard geodesic equation:
\begin{equation}
    \frac{\tilde{D}}{ds}\left( \frac{dx^k}{ds} \right) = \frac{d^2 x^k}{ds^2} + \tilde{\Gamma}^k_{ij} \frac{dx^i}{ds} \frac{dx^j}{ds} = 0.
\end{equation}

\begin{figure}[t]
    \centering
    \includegraphics[width=\linewidth]{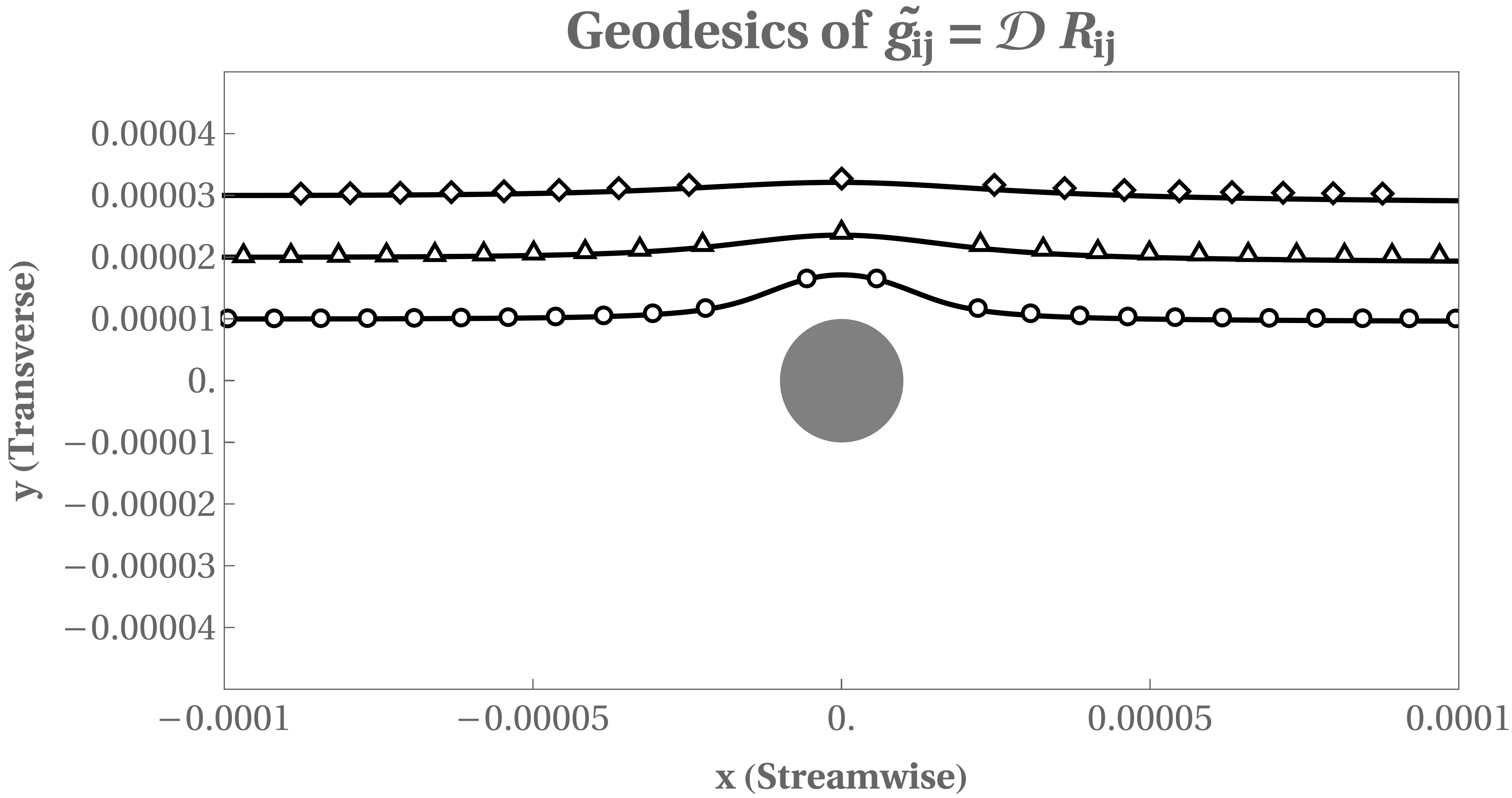}
    \caption{Geodesics of the metric $\tilde{g}_{ij} = \mathcal{D}\,\tilde{R}_{ij}$ (solid lines) compared with analytical Risbud-Drazer trajectories \cite{RisbudDrazer2013} (open symbols). Parameters correspond to a microparticle in water: $a = 10~\mu$m each, moving in water ($\mu=10^{-3}$ Pa$\cdot$s, $\rho_f=1000$ kg/m$^3$, particle density $2000$ kg/m$^3$). Impact parameters are $\{a, 2a, 3a\}$ from bottom to top.}
    \label{fig:gTilIsTrajwRD}
\end{figure}

Equation~\eqref{eqn:timedilationfinal} shows that the affine parameter of the geodesic flow is the cumulative energy dissipated into the fluid:
\begin{equation}
    s(t) = \int_0^t \mathcal{D}(\mathbf{x}(\tau)) \, d\tau = \Delta E_{\text{diss}}.
\end{equation}
Thus, physical particle trajectories under a constant external force correspond to stationary paths of cumulative energy dissipation in the metric $\tilde{g}_{ij} = \mathcal{D}\tilde{R}_{ij}$. While the Helmholtz theorem applies to the velocity field at each instant, this variational formulation characterizes the continuous particle trajectory. Figure~\ref{fig:gTilIsTrajwRD} illustrates this correspondence for the two-sphere system, contrasting the physical trajectories with the unscaled resistance geodesics in Figure~\ref{fig:gIsRTraj}.

\subsection{Relation Between First-Order Dynamics and Geodesic Flow}
Although Stokesian particle motion is governed by the local first-order kinematic relation $\mathbf{U} = \tilde{\mathbf{R}}^{-1}(\mathbf{x})\cdot\mathbf{F}$, the resulting trajectories satisfy a second-order geodesic equation. This behavior is analogous to geometric optics: light rays satisfy Fermat's principle of stationary travel time while locally following Snell's law at every point. In the Stokesian problem, the resistance tensor $\tilde{\mathbf{R}}(\mathbf{x})$ is determined by the global boundaries of the fluid domain, establishing a metric space in which local mobility steps integrate into geodesics.

In this variational context, the term ``admissible paths'' refers to virtual curves $\gamma(t)$ connecting endpoints $\mathbf{x}_A$ and $\mathbf{x}_B$ that do not necessarily satisfy the mobility relation $\dot{\gamma} = \tilde{\mathbf{R}}^{-1}\cdot\mathbf{F}$. Extremizing the functional $S[\gamma] = \int \sqrt{\tilde{g}_{ij} \dot{x}^i \dot{x}^j} dt$ across these virtual curves yields the second-order geodesic equations on $(\mathcal{M}, \tilde{\mathbf{g}})$. The physical trajectories prescribed by the first-order mobility equation form an invariant family (a geodesic congruence) of this manifold, analogous to how rays in optics form a congruence associated with an eikonal wave front.

\section{Application to a two-sphere system}\label{sec:two-sphere-system}
Risbud and Drazer \cite{RisbudDrazer2013} derived analytical equations describing the trajectory of a sphere moving past another fixed sphere in an unbounded Stokesian fluid:
\begin{equation}
b_{in} = y\exp{H(r)},
\label{eqn:spheretrajectory}
\end{equation}
where $b_{in}$ is the asymptotic impact parameter, $y$ is the transverse coordinate, and $r$ is the radial separation. The function $H(r)$ is:
\begin{equation}
H(r) = \displaystyle\int_r^\infty \frac{R_B(s) - R_A(s)}{sR_B(s)} ds.
\end{equation}
Given the axisymmetry of the two-body geometry, the modified effective resistance tensor decomposes into longitudinal and transverse components \cite{RisbudDrazer2013}:
\begin{equation}
    \tilde{\mathbf{R}} = R_A(r) \mathbf{\hat{r}}\mathbf{\hat{r}} + R_B(r)\left(\boldsymbol{\delta} - \mathbf{\hat{r}}\mathbf{\hat{r}}\right),
\label{eqn:twosphresis}
\end{equation}
where $\mathbf{\hat{r}}$ is the unit vector along the line of centers.

\subsection{Longitudinal and Transverse Resistance}
The functions $R_A(r)$ and $R_B(r)$ account for hydrodynamic interactions between the spheres \cite{JeffreyOnishi1984, KimKarrila, RisbudDrazer2013}:
\begin{enumerate}
    \item \textbf{Longitudinal Resistance $R_A(r)$:} Along the line of centers $\mathbf{\hat{r}}$, symmetry precludes translation-rotation coupling ($\mathbf{R}_{\mathbf{F\omega}}\cdot\mathbf{\hat{r}} = \mathbf{0}$). Motion along $\mathbf{\hat{r}}$ corresponds to squeeze flow out of the narrowing gap, giving $R_A(r) = (\mathbf{R}_{\mathbf{FU}})_r(r) = 1/x_{11}^a(r)$, where $x_{11}^a(r)$ is the longitudinal mobility function. In the lubrication limit ($r \to 2a$, with $\epsilon = r/a - 2 \ll 1$), $x_{11}^a(r) \sim \mathcal{O}(\epsilon)$, so the longitudinal resistance diverges as $R_A \sim \mathcal{O}(1/\epsilon)$.
    \item \textbf{Transverse Resistance $R_B(r)$:} Perpendicular to $\mathbf{\hat{r}}$, the effective transverse resistance for a torque-free particle is $R_B(r) = (\mathbf{R}_{\mathbf{FU}})_\perp - (\mathbf{R}_{\mathbf{F\omega}})_\perp^2 / (\mathbf{R}_{\mathbf{T\omega}})_\perp = 1/y_{11}^a(r)$, where $y_{11}^a(r)$ is the torque-free transverse mobility \cite{JeffreyOnishi1984, RisbudDrazer2013}. In lubrication, $y_{11}^a(r) \sim \mathcal{O}(1/\ln(1/\epsilon))$, so $R_B(r)$ grows logarithmically.
\end{enumerate}

\subsection{Anisotropic Curvature and Trajectory Deflection}
In our geometric framework, Eq.~\eqref{eqn:spheretrajectory} is the exact first integral of the geodesic equation on the manifold $(\mathcal{M}, \tilde{\mathbf{g}} = \mathcal{D}\tilde{\mathbf{R}})$. Figure~\ref{fig:gTilIsTrajwRD} demonstrates the agreement between the geodesics of $\tilde{\mathbf{g}}$ and the analytical trajectories of Risbud and Drazer \cite{RisbudDrazer2013}.

The function $H(r)$ quantifies the anisotropic curvature of this dissipation manifold. In the hypothetical isotropic limit where $R_A(r) = R_B(r) = R(r)$, the mobility is isotropic ($\mathbf{U} \parallel \mathbf{F}$). The dissipation rate is $\mathcal{D} = F^2/R(r)$, and the metric simplifies to:
\begin{equation}
    \tilde{\mathbf{g}} = \left(\frac{F^2}{R}\right) R \boldsymbol{\delta} = F^2 \boldsymbol{\delta}.
\end{equation}
Because the metric is proportional to the Euclidean identity tensor with a constant factor, the Christoffel symbols vanish, and the geodesics are straight lines ($y = b_{in}$).

In reality, hydrodynamic interactions break isotropy ($R_A > R_B$). The dimensionless quantity:
\begin{equation}
    \Delta n(r) \equiv \frac{R_B(r) - R_A(r)}{R_B(r)} < 0
\end{equation}
acts as an anisotropic index gradient. The obstacle creates a region of elevated longitudinal resistance, deflecting the particle laterally, while the conformal factor $\mathcal{D}$ ensures that the trajectory bends symmetrically downstream.

\section{Discussion and Conclusions}
We have presented a differential geometric framework that identifies the metric $\tilde{\mathbf{g}} = \mathcal{D}(\mathbf{x}) \tilde{\mathbf{R}}(\mathbf{x})$ as the geometry of motion for a torque-free particle driven by a constant external force in Stokes flow. Physical trajectories are exact geodesics of this manifold, with the affine parameter corresponding to cumulative viscous dissipation.

\subsection{Relationship to Geometric Swimming}
This framework is complementary to the gauge-theoretic description of low-Reynolds-number self-propulsion established by Shapere and Wilczek \cite{ShapereWilczek1989a, ShapereWilczek1989b} and the relativistic treatment of swimming in curved spacetime by Wisdom \cite{Wisdom2003}. In geometric swimming, hydrodynamics is encoded in a gauge connection over the manifold of body shapes; cyclic shape changes produce net translation through geometric holonomy, with optimal strokes corresponding to geodesics on shape space.

The primary distinction is the driving mechanism: Shapere and Wilczek address active matter undergoing internal deformations without net external force, whereas the present framework addresses passive rigid particles driven through physical space by external forces. Both frameworks, however, share a common physical basis in the minimization of viscous dissipation. The efficiency metric on shape space is mathematically dual to the resistance metric on physical space, with $\tilde{\mathbf{g}}$ serving as the passive-transport counterpart to the active-swimming gauge potential.

\subsection{Optimal Paths: Controlled vs. Force-Driven Motion}
These results outline two classes of optimal paths in Stokesian systems:
\begin{enumerate}
    \item \textbf{Resistance Geodesics ($\tilde{\mathbf{R}}$):} Paths that minimize the integrated resistance $\int \tilde{R}_{ij} \dot{x}^i \dot{x}^j dt$ between two points. These represent optimal paths for actively controlled particles, such as those steered by optical tweezers or magnetic fields, where a controller adjusts the applied force to minimize total work done against the fluid.
    \item \textbf{Dissipative Geodesics ($\tilde{\mathbf{g}}$):} Natural trajectories of particles driven by constant, unidirectional external forces (e.g., gravity or uniform electric fields). Here, the particle balances the resistance gradient against the constant force, following a path of stationary dissipative action.
\end{enumerate}
The drift term in Eq.~\eqref{eqn:geodesic_deviation}, $\frac{DU^i}{dt} = \frac{1}{2}g^{im}\partial_m\mathcal{D}$, quantifies the deviation of the force-driven trajectory from the path of minimum resistance due to the constant force orientation.

\subsection{Broader Implications}
This geometric framework suggests several broader directions:

\textbf{Trajectory-level variational principles:} Classical variational formulations of Stokes flow \cite{Helmholtz1868, KellerRubenfeldMolyneux1967} apply to the instantaneous velocity field. The present framework extends variational methods to particle trajectories in Stokes flow, relating local dissipation to path geometry.

\textbf{Particle separation in microfluidics:} In microfluidic sorting, including deterministic lateral displacement (DLD) pillar arrays and porous media, separation relies on particle-obstacle hydrodynamic interactions. Because particles of different sizes, shapes, or surface properties have distinct resistance tensors $\bm{\mathcal{R}}(\mathbf{x})$, they experience different metric fields $\tilde{\mathbf{g}}(\mathbf{x})$ and follow different geodesics. Mapping these metrics could aid in designing obstacle arrangements that separate particles based on their hydrodynamic signatures.

\textbf{Many-body suspensions:} Although we focused on a single particle navigating fixed obstacles, the formulation extends in principle to multi-particle systems. For an $N$-particle suspension, the grand resistance tensor defines a Riemannian metric on a $6N$-dimensional configuration manifold. The evolution of the suspension under external driving corresponds to a trajectory on this manifold, where curvature metrics could help characterize suspension rheology and collective transport.

\textbf{Stochastic transport:} In colloidal regimes where thermal fluctuations are significant, deterministic trajectories are replaced by probability distributions. Coupling the metric $\tilde{\mathbf{g}}$ with the Onsager-Machlup functional may clarify whether the most probable path of a Brownian particle coincides with the geodesic of $\tilde{\mathbf{g}}$ or experiences noise-induced drift from manifold curvature.

\begin{acknowledgments}
The author is grateful to Prof. German Drazer of Rutgers University, Prof. Ravindra Kulkarni of Bhaskaracharya Pratishthana (Pune), and Prof. Rahul Kashyap of IIT Bombay for enlightening discussions.
\end{acknowledgments}

\bibliography{apssamp}

\end{document}